\documentclass[12pt]{article}
\usepackage{color}
\usepackage{setspace}
\usepackage[margin=1in,lines=25]{geometry}
\usepackage{amsmath,amsthm,amssymb,bm}
\usepackage{graphicx}
\usepackage{algorithm}
\usepackage{algorithmic}
\usepackage{multirow}
\usepackage{longtable}
\usepackage{physics}
\usepackage{rotating}
\usepackage{verbatim}
\RequirePackage[colorlinks,citecolor=blue,linkcolor=blue]{hyperref}
\usepackage{natbib}

\DeclareMathOperator*{\argmin}{arg\,min}
\makeatletter
\def\maxwidth{ %
  \ifdim\Gin@nat@width>\linewidth
    \linewidth
  \else
    \Gin@nat@width
  \fi
}
\makeatother

\usepackage{color}
\definecolor{fgcolor}{rgb}{0, 0, 0}
\newcommand{\hlnum}[1]{\textcolor[rgb]{0.533,0,0.133}{#1}}%
\newcommand{\hlopt}[1]{\textcolor[rgb]{0,0,0}{\textbf{#1}}}%
\newcommand{\hlstd}[1]{\textcolor[rgb]{0,0,0}{#1}}%
\newcommand{\hlkwa}[1]{\textcolor[rgb]{0.4,0.067,0.067}{\textbf{#1}}}%
\newcommand{\hlkwb}[1]{\textcolor[rgb]{0,0,0.4}{\textbf{#1}}}%
\newcommand{\hlkwc}[1]{\textcolor[rgb]{0,0,0.4}{#1}}%
\newcommand{\hlkwd}[1]{\textcolor[rgb]{0,0.267,0.4}{#1}}%

\usepackage{framed}
\makeatletter
\newenvironment{kframe}{%
 \def\at@end@of@kframe{}%
 \ifinner\ifhmode%
  \def\at@end@of@kframe{\end{minipage}}%
  \begin{minipage}{\columnwidth}%
 \fi\fi%
 \def\FrameCommand##1{\hskip\@totalleftmargin \hskip-\fboxsep
 \colorbox{shadecolor}{##1}\hskip-\fboxsep
     \hskip-\linewidth \hskip-\@totalleftmargin \hskip\columnwidth}%
 \MakeFramed {\advance\hsize-\width
   \@totalleftmargin\z@ \linewidth\hsize
   \@setminipage}}%
 {\par\unskip\endMakeFramed%
 \at@end@of@kframe}
\makeatother

\definecolor{shadecolor}{rgb}{.97, .97, .97}
\definecolor{messagecolor}{rgb}{0, 0, 0}
\definecolor{warningcolor}{rgb}{1, 0, 1}
\definecolor{errorcolor}{rgb}{1, 0, 0}
\newenvironment{knitrout}{}{} 

\usepackage{alltt}
\usepackage{natbib}
\usepackage{amsfonts,amssymb,graphics,epsfig,verbatim,bm}
\usepackage{latexsym,amsmath,url,amsbsy}
\usepackage{graphicx, cleveref}
\usepackage{booktabs,setspace}
\usepackage{caption}
\usepackage{multirow}

\usepackage{color}
\newcommand{\blue}[1]{{\textcolor{black}{#1}}}

\newcommand{\code}[1]{\blue{\texttt{#1}}}

\renewcommand{\hat}{\widehat}
\renewcommand{\bar}{\overline}
\renewcommand{\tilde}{\widetilde}

\begin{document}

\title{Heterogeneous Regression Models for Clusters of Spatial Dependent Data}

\author{Zhihua Ma~~~~Yishu Xue~\thanks{{yishu.xue@uconn.edu}} ~~~~ Guanyu
                Hu}
\date{}
\maketitle 

\begin{abstract}
In economic development, there are often regions that share similar economic
characteristics, and economic models on such regions tend to have similar
covariate effects. In this paper, we propose a Bayesian clustered regression for
spatially dependent data in order to detect clusters in the covariate effects.
Our proposed method is based on the Dirichlet process which provides  a
probabilistic framework for simultaneous inference of the number of clusters and
the clustering configurations.
The usage of our method is illustrated both in simulation studies and an
application to a housing cost dataset of Georgia.
\bigskip

\noindent 
\textit{keywords:} Clustered Coefficients Regression, Dirichlet process, 
  MCMC,  Spatial Random Effects

\end{abstract}

\maketitle
\newpage 

\section{Introduction}\label{sec:intro}
Spatial regression models have been widely applied in many different fields such
as environmental science \citep[][]{hu2018stat}, biological science
\citep[][]{zhang2011bayesian}, and econometrics
\citep[][]{brunsdon1996geographically} to explore the relation between a
response variable and a set of predictors over a region. One of the most
important tasks for a spatial regression model is to capture the spatial
dependent structure between a response variable and a set of covariates.
\citet{cressie1992statistics} proposed a spatial regression model with Gaussian
process, where the spatial random effects are accounted for only
by the intercepts.
\cite{brunsdon1996geographically} proposed a geographically weighted
regression (GWR) model which assumed the existence of a spatially-dependent
parameter surface, and used weighted local linear regression to estimate such
parameter surface. An application of GWR in analyzing the impact of
socio-economic factors on treated prevalence for mental disorders in
Barcelona is presented in
\cite{peres2015applying}.
The idea of GWR has been subsequently extended to the Cox
model framework by \cite{xue2018geographically}.
In addition to mean-based regression, \cite{chasco2015heterogeneity}
used a spatial quantile regression technique to identify heterogeneities.
From the Bayesian perspective,
\cite{gelfand2003spatial} incorporated Gaussian process to regression
coefficients to build a model with spatially varying coefficients.
\cite{autantbernard2019heterogeneous} used a Bayesian
heterogeneous spatial autoregressive model that allows for 
spatial variation variations in intercepts, covariate effects,
and noise variances to study the knowledge
production functions of different regions in order
to set up their regional innovation strategy. 
The
aforementioned works, however, all assumed that each location has its own set of
regression parameters, which sometimes leads to excessive numbers of
parameters, and subsequently overfitting. Cluster effects over the space of
interest has not been taken into account.

Detection of heterogeneous covariate effects in many different fields, such as
real estate applications, spatial econometrics, and environmental science are
becoming of increasing research interest. For example, administrative divisions
in a country, such as regions, provinces, states, or territories, often have
different economic statuses and development patterns. More advanced divisions
and less developed divisions could be put into separate clusters and analyzed.
Such clustering information is of great interest to regional economics
researchers. One of the most popular methods for spatial cluster detection is
the scan statistic method \citep{kulldorff1995spatial}, where a scan statistic
is constructed via a likelihood ratio statistic to test the potential clusters.
The usage of spatial scan statistics has been extended to studies of disease
mapping, 
crime, 
and public health.
Similar endeavor has also been made under the Bayesian and nonparametric Bayesian
frameworks in pursuit of spatial homogeneity.
\cite{li2015bayesian} used
nonparametric Bayesian method to detect cluster boundaries for areal data.
Noticing that
traditional methods may not work well with spatial missing data,
\cite{panzera2016bayesian} proposed using multiple imputation
together with the Bayesian Interpolation method to analyze spatially
clustered missing data, which addresses
both spatial univariate and multivariate problems.
%

Most of the aforementioned frequentist and Bayesian approaches
mainly focus on estimating cluster configurations of spatial response.
Spatially varying patterns in the relationship between a set of covariates
and the response is also an important topic that needs to be studied.
\cite{bille2017twostep} used a two-step approach where in the first step,
spatial regimes of spatially varying parameters are identified, and in the
second step estimated.
Recently,
methods for cluster detection of spatial regression coefficients have been
proposed to detect the homogeneity of the covariates effects among subareas.
\citet{li2019spatial} incorporated spatial
neighborhood information in a penalized approach to detect spatially clustered
patterns in the regression coefficients.

Under the Bayesian framework, \cite{lawson2014prior} explored the usage of
multinomial priors in modeling clustered coefficients in the accelerated failure
time model for survival data. As discussed by \cite{lawson2014prior}, to infer
the grouping level,
complicated search algorithms in variable dimensional parameter space are
needed, such as the reversible jump Markov chain Monte Carlo (MCMC) algorithm
of \cite{green1995reversible}, which assigns a prior on the
number of clusters, and this number is updated at each iteration of an
MCMC chain.
Such algorithms are difficult to implement and automate, and are known to suffer
from lack of scalability and mixing issues. Nonparametric Bayesian approaches,
such as the Dirichlet process mixture model
\citep[DPMM;][]{ferguson1973bayesian}, offer choices to allow for uncertainty in the
number of clusters, and provide an integrated probabilistic framework
under which the number of clusters, the clustering configuration, and regression
coefficients are simultaneously estimated.

In this paper, we propose a Bayesian spatial clustered linear regression model
with \blue{Dirichlet process} \citep[DP;][]{ferguson1973bayesian} prior, which
considers spatially dependent structure and clusters the covariate effects
simultaneously. In addition, implementation of our proposed methods based on
\textbf{nimble} \citep[][]{de2017programming}, a relatively new and powerful
\textsf{R} package, is discussed. The model diagnostic technique, logarithm of
the pseudo-marginal likelihood \citep[LPML;][]{ibrahim2013bayesian}, is
introduced to assess the fitness of our proposed model. Our proposed Bayesian
approach reveals interesting features of the state-level data of Georgia.

The remainder of the paper is organized as follows. In Section~\ref{sec:method},
we develop a spatial clustered linear regression model with DP prior. In
Section~\ref{sec:bayes_comp}, a MCMC sampling algorithm based on
\textbf{nimble}
and post MCMC inference are discussed. Extensive simulation studies are carried
out in the next section. For illustration, our proposed methodology is
applied to Georgia housing cost dataset in Section~\ref{sec:real_data}.
Finally, we conclude this paper with a brief discussion.

\section{Methodology}\label{sec:method}

In this section, a Bayesian spatial clustered linear model using \blue{DPMM} is
proposed for coefficient grouping in spatially dependent data. Based on the
spatial regression model, spatially-varying coefficients are assigned with a
nonparametric \blue{DP} prior to achieve the goal of grouping.

\subsection{Spatial Regression Model}
The basic geostatistical model \citep[][]{gelfand2016spatial} for spatially
dependent response at locations $\bm{s} = (s_1,\ldots, s_n)$ is denoted
by
\begin{equation}
\label{eq:spatial_regression}
\bm{Y}=\bm{X}\bm{\beta}+\bm{w}+\bm{\epsilon},
\end{equation}
\blue{where $\bm{Y} = (Y(s_1),\ldots, Y(s_n))$ the $n$-dimensional vector of
responses observed at the~$n$ different locations, $\bm{X} = \begin{pmatrix} 
    X(s_1)^\top \\
    \ldots \\
    X(s_n)^\top
\end{pmatrix}$ is the~$n\times p$ matrix of covariates, $\bm{w} = (w(s_1),\ldots,
w(s_n))$ is the vector of spatial random effects, which is assumed to
follow a stationary Gaussian process whose covariance structure often depends on
the geographical locations, and
$\bm{\epsilon}\sim\mbox{MVN}(\bm{0}, \sigma^2_y \bm{I})$} adds the nugget effect
\citep[\blue{see} e.g., Chapter~6 of][]{carlin2014hierarchical}, which is
usually
a vector of  white noise, \blue{with MVN denoting the multivariate normal distribution.}
 Oftentimes,~$\bm{w}(\bm{s})$
is assumed to also follow a MVN.
The above spatial regression model can also be rewritten as
\begin{equation*}
\label{eq:spatial-regression-2}
\begin{split}
\bm{Y} \mid \bm{\beta}, \bm{w}, \sigma_y^2 &\sim
\mbox{MVN}(\bm{X}\bm{\beta} + \bm{w},
\sigma_y^2\bm{I}),\\
\bm{w} &\sim \mbox{MVN}(\bm{0},\bm{\Sigma}_W),
\end{split}
\end{equation*}
where
~$\bm{\Sigma}_W$ \blue{is the covariance matrix of the spatial random
effect vector $\bm{w}(\bm{s})$}, and~$\bm{I}$ denotes the identity matrix.
Conditional on $\bm{X}$ and $\bm{w}$, entries in
$\bm{Y}$ are independent.
Conventionally, the
covariance matrix is given as $\bm{\Sigma}_W = \sigma_w^2 \bm{H}$,
where $\bm{H}$ is a matrix
constructed using the great circle distance matrix, denoted as GCD,
between different locations, i.e.,
\begin{equation*}
    \mbox{GCD} = \mbox{GCD}(i,j)_{i,j\in\{1,\ldots,n\}} = 
    \mbox{great circle distance between locations $s_i$ and $s_j$},
\end{equation*}
and~$\sigma_w^2$ is a scalar. There are
three common weighting schemes for 
\blue{defining}~$\bm{H}$, including:
\begin{equation}\label{eq:weightingscheme}
\begin{split}
\text{the unity scheme}: & ~~\bm{H} = \bm{I}_{n\times n} \\
\text{the exponential scheme}: & ~~\bm{H}(\phi)  = \bm{H}(i,j)_{i,j\in\{1,\ldots,n\}} =
\left[\exp(-\mbox{GCD}(i,j) / \phi)\right]_{n\times n}, \\
\text{the Gaussian scheme}: & ~~\bm{H} (\phi) = \bm{H}(i,j)_{i,j\in\{1,\ldots,n\}} =
\left[\exp(-(\mbox{GCD}(i,j)/\phi)^2)\right]_{n\times n},
\end{split}
\end{equation}
where $\phi$ is a tuning parameter that controls the spatial
correlation. Larger
value of $\phi$ indicates stronger correlation.

Another spatial regression model is the spatially varying coefficients model
\citep[][]{gelfand2003spatial}:
\begin{eqnarray}
	Y(s)=\bm{X}(s)^\top\tilde{\bm{\beta}}(s)+\epsilon(s),
	\label{svc model}
\end{eqnarray}
where $X(s)$ is $p \times 1$ covariate vector at \blue{a certain} location $s$,
and~$\tilde{\bm{\beta}}(s)$ is assumed to follow a~$p$-variate spatial process
model. If we have observations $(Y(s_i),\bm{X}(s_i))$ for $i=1,\ldots, n$, they
can be written into 
\begin{equation*}
  \bm{Y} = \bm{X}^\top \tilde{\bm{\beta}} + \bm{\epsilon},
\end{equation*} 
where $\bm{Y}=(Y(s_1), \ldots, Y(s_n))^\top$, $\bm{X}^\top$ is an $n\times (np)$
block diagonal matrix which has the row vector $\bm{X}^\top(s_i)$ as its $i$-th
diagonal entry, $\tilde{\bm{\beta}} = (\tilde{\bm{\beta}}(s_1)^\top, \ldots,
\tilde{\bm{\beta}}(s_n)^\top)^\top$, and~$\bm{\epsilon}\sim
\mbox{MVN}(\bm{0},\sigma^2\bm{I})$. \citet{gelfand2003spatial} proposed the following
hierarchical model:
\begin{equation}
\begin{split}
&\bm{Y} \mid \tilde{\bm{\beta}},\sigma^2 \sim
\mbox{MVN}(\bm{X}^\top\tilde{\bm{\beta}},\sigma^2\bm{I})\\
&{\tilde{\bm{\beta}}}\mid \bm{\mu}_{\bm{\beta}},\bm{T} \sim
\mbox{MVN}(1_{n\times
1}\otimes
\bm{\mu}_{\bm{\beta}},\bm{H}(\phi)\otimes \bm{T})
\end{split}
\label{hierarchical model}
\end{equation}
where $\bm{\mu}_{\bm{\beta}}$ is a $p \times 1$ vector,  $\bm{H}(\phi)$ is a
$n\times n$ matrix measuring spatial correlations between the $n$ observed
locations, $\bm{T}$ is a $p \times p$ covariance matrix associated with an
observation vector at any spatial location, and $\otimes$ denotes the Kronecker
product. \blue{In model \eqref{eq:spatial_regression}, $\bm{\beta}$ is constant
over space, which means the covariate effects remain the same over all
locations; model~\eqref{hierarchical model} allows for different covariate
effects over locations, but restrict the covariate effects to be determined by
distances between pairs of locations as in~\eqref{eq:weightingscheme}.}


For
many spatial economics data, however, some regions will share similar covariate
effects regardless of their geographical distance. Taking China as an example,
Beijing tends to have similar economic development pattern with Shanghai or
Jiangsu \citep[][]{ma2019nonparametric} . The models in
\eqref{eq:spatial_regression} and \eqref{hierarchical model}, however, do not
take into account such inherent similarities in spatially dependent data.

\subsection{Spatial Regression with Dirichlet Process Mixture Prior}

Within the Bayesian framework, coefficient clustering can be accomplished using
a Dirichlet process mixture model (DPMM) by nonparametrically linking the
spatial response variable to covariates through cluster membership.
Formally, a probability measure $G$ following
a DP with a concentration
parameter~$\alpha$ and a base
distribution~$G_0$  is denoted by~$G \sim
\text{DP}(\alpha, G_0)$~if
\begin{equation}
\label{eq:DP}
(G(A_1), \cdots, G(A_r)) \sim \text{Dirichlet}(\alpha G_0(A_1), \cdots, \alpha
G_0(A_r)),
\end{equation}
where $(A_1, \cdots, A_r)$ are finite measurable partitions of the space
$\Omega$. Several different formulations can be used for determining the DP. In
this
work, we use the stick-breaking construction proposed by
\cite{sethuraman1994constructive} for DP realization, which is given as
\begin{gather*}
\theta_c \sim G_0, ~~ G = \sum_{c=1}^{\infty} \pi_c
\delta_{\theta_c}(\cdot),\\
\pi_1 = V_1, ~~ \pi_c = V_c \prod_{\ell<c}(1-V_\ell),\\
V_c \overset{\mbox{ind}}{\sim} \text{Beta}(1, \alpha),
\end{gather*} 
where $\theta_c$ is the $c$-th vector consisting of the possible values
of the
parameters of $G_0$, $\delta_{\theta_c}(\cdot)$ denotes a discrete probability
measure concentrated at $\theta_c$ and is a notation short for
$\delta(\theta = \theta_c)$, $\pi_c$ is the random probability
weight between 0 and 1, and $\overset{\mbox{ind}}{\sim}$ indicates
i.i.d..

For a DPMM, the observed data $y_i$ (for $i=1,\ldots, n$) follow an infinite
mixture distribution, where a vector of latent allocation variables
$\mathcal{Z}$ is introduced to enable explicit characterization of the
clustering. Let $\mathcal{Z}_{n, k} = \big\{(z_1, \ldots, z_n) : z_i \in \{1,
\ldots, k\}, 1 \le i \le n \big\}$ denote all possible clusterings of $n$
observations into $k$ clusters, where $z_i = c \in \{1, \ldots, k\}$ denotes the
cluster assignment of the $i$th observation. Note that although
theoretically $c$ can
go to infinity, in practice, with $n$ observations in total, $k$ is capped
at~$n$,
which can only happen when each observation is assigned to its own cluster.
The DPMM can be written as 
\begin{equation}
\begin{split}
y_i\mid\mathcal{Z}, \theta &\sim f(y_i \mid \theta_{z_i}),\\
\theta_c &\sim G_0,\\
P(z_i = c \mid &\bm{\pi}) = \pi_c,\\
\pi_1 = V_1, ~~ \pi_c = V_c
\prod_{\ell<c}(&1-V_\ell),\hspace{0.1cm}
V_c \overset{\mbox{ind}}{\sim} \text{Beta}(1, \alpha),
\end{split}
\end{equation}
where $\bm{\pi} = (\pi_1,\ldots, \pi_c, \ldots)$. Adapting the DPMM to the
spatial regression setting, we focus on the clustering of spatially-varying
coefficients $\bm{\beta}(s) = (\bm{\beta}^\top(s_1), \cdots,
\bm{\beta}^\top(s_n))^\top$, where $\bm{\beta}(s_i)$ is the $p$-dimensional
coefficient vector for location $s_i$. In our setting, we assume that the $n$
parameter vectors can be clustered into $k$ groups, i.e., $\bm{\beta}(s_i) =
\bm{\beta}_{z_i}\in\{\bm{\beta}_1,\ldots, \bm{\beta}_k\}$, then the model can be
written as
\begin{align}
y(s_i) \mid \bm{\beta}_{z_i}\blue{,}  w(s_i),  \sigma_y^2 &\sim
N(\bm{X}(s_i)^\top\bm{\beta}(s_{z_i})+w(s_i), \sigma_y^2),\label{eq:ydist} \\ 
\bm{w}(\bm{s}) &\sim \mbox{MVN}(\bm{0},\bm{\Sigma}_W),\label{eq:Wdist}\\
\bm{\beta}_{z_i} &\stackrel{\text{ind}} \sim \mbox{MVN}(\bm{\mu}, \bm{\Sigma}),\\
P(&z_i = c \mid \bm{\pi}) = \pi_c,\label{eq:pzi}\\
\pi_1 = V_1, ~~ \pi_c = V_c &\prod_{\ell<c}(1-V_\ell),
~~ V_c \overset{\mbox{ind}}{\sim} \text{Beta}(1, \alpha)\label{eq:breaking}.
\end{align}

\section{Bayesian Inference}\label{sec:bayes_comp}

MCMC is used to draw samples from the posterior distributions of the model
parameters. In this section we present the sampling scheme, the posterior
inference of cluster belongings, and measurements to evaluate the estimation
performance and clustering accuracy.

\subsection{The MCMC Sampling Scheme}

We present the main \texttt{R} function written using the \textbf{nimble}
package \citep[][]{de2017programming}. The model is wrapped in a
\texttt{nimbleCode()} function. For ease of exposition, we break it into
separate snippets. The full code is available on GitHub with an example
implementation. (link removed for blinding purposes, documentation submitted
separately)

Define \code{S} as the number of locations. The following code represent
Equation~\eqref{eq:ydist}. At each location, \code{y[i]} has a normal
distribution with \code{mu\_y[i]} and precision \code{tau\_y}, which is
equivalent to $1/\sigma_y^2$. A Gamma(1,1) prior is given to \code{tau\_y}. The
coefficient vector for location \code{i}, \code{b[i, 1:6]}, equals the
coefficient vector estimated for the cluster it belongs to, represented by
\code{latent[i]}, which follows a multinomial distribution with probability
vector \code{zlatent[1:M]}, where \code{M} denotes the number of potential
clusters.

\begin{knitrout}
\definecolor{shadecolor}{rgb}{0.965, 0.965, 0.965}\color{fgcolor}\begin{kframe}
\begin{alltt}
SLMMCode <- \hlkwd{nimbleCode}(\{
  \hlkwd{for} (i in 1:S) \{
    y[i] ~ \hlkwd{dnorm}(mu_y[i], tau = tau_y)
    mu_y[i] <- b[i, 1] * x1[i] + b[i, 2] * x2[i] +
      b[i, 3] * x3[i] + b[i, 4] * x4[i] + b[i, 5] * x5[i] +
      b[i, 6] * x6[i] + W[i]
    
    b[i, 1:6] <- bm[latent[i], 1:6]

    latent[i] ~ \hlkwd{dcat}(zlatent[1:M])
  \}
\end{alltt}
\end{kframe}
\end{knitrout}
{\setlength{\parindent}{0cm}
The following code represent Equation~\eqref{eq:Wdist}. \code{H} represents the
matrix $\Sigma_W$, where \code{phi} is \blue{the} tuning parameter $\phi$ in
the exponential scheme in \eqref{eq:weightingscheme}
that controls spatial
correlation. The random effects at locations 1 to \code{S} follow a multivariate
normal distribution with \code{mu\_w[1:S]} and precision matrix, which equals
the product of $\sigma_w^2$, \code{tau\_w}, and the inverse of \code{H}.
\code{H} is defined as a function of a certain distance matrix
\code{Dist}, which is passed into the function later as an argument
of \code{SLMMConsts}. Here, the function is chosen to be the exponential
scheme.
The
prior distribution of the bandwidth \code{phi} is specified to be a uniform
distribution from 0 to a certain upper limit, denoted by \code{D}. The prior of
\code{tau\_w} is set to Gamma(1,1).}
\begin{knitrout}
\definecolor{shadecolor}{rgb}{0.965, 0.965, 0.965}\color{fgcolor}\begin{kframe}
\begin{alltt}
  \hlkwa{for} \hlstd{(j} \hlkwa{in} \hlnum{1}\hlopt{:}\hlstd{S) \{}
    \hlkwa{for} \hlstd{(k} \hlkwa{in} \hlnum{1}\hlopt{:}\hlstd{S) \{}
      \hlstd{H[j, k]} \hlkwb{<-} \hlkwd{exp}\hlstd{(}\hlopt{-}\hlstd{Dist[j, k]}\hlopt{/}\hlstd{phi)}
    \hlstd{\}}
  \hlstd{\}}

  \hlstd{W[}\hlnum{1}\hlopt{:}\hlstd{S]} \hlopt{~} \hlkwd{dmnorm}\hlstd{(mu_w[}\hlnum{1}\hlopt{:}\hlstd{S],} \hlkwc{prec} \hlstd{= prec_W[}\hlnum{1}\hlopt{:}\hlstd{S,} \hlnum{1}\hlopt{:}\hlstd{S])}
  \hlstd{prec_W[}\hlnum{1}\hlopt{:}\hlstd{S,} \hlnum{1}\hlopt{:}\hlstd{S]} \hlkwb{<-} \hlstd{tau_w} \hlopt{*} \hlkwd{inverse}\hlstd{(H[}\hlnum{1}\hlopt{:}\hlstd{S,} \hlnum{1}\hlopt{:}\hlstd{S])}

  \hlstd{phi} \hlopt{~} \hlkwd{dunif}\hlstd{(}\hlnum{0}\hlstd{, D)}
  \hlstd{tau_w} \hlopt{~} \hlkwd{dgamma}\hlstd{(}\hlnum{1}\hlstd{,} \hlnum{1}\hlstd{)}

  \hlstd{mu_w[}\hlnum{1}\hlopt{:}\hlstd{S]} \hlkwb{<-} \hlkwd{rep}\hlstd{(}\hlnum{0}\hlstd{, S)}
\end{alltt}
\end{kframe}
\end{knitrout}
{\setlength{\parindent}{0cm}
The distribution of $\bm{\beta}$ for each location $s_i$ is defined next. They
each come from a multivariate normal distribution with mean \code{mu\_bm} and
covariance matrix \code{var\_bm}, which is a diagonal matrix with all diagonal
entries being \code{1/tau\_bm}. The inverse variance term, \code{tau\_bm},
is
again given a Gamma(1,1) prior, and the entries in the mean vector are all given
independent standard normal priors.
}
\begin{knitrout}
\definecolor{shadecolor}{rgb}{0.965, 0.965, 0.965}\color{fgcolor}\begin{kframe}
\begin{alltt}
  \hlkwa{for} \hlstd{(k} \hlkwa{in} \hlnum{1}\hlopt{:}\hlstd{M) \{}
    \hlstd{bm[k,} \hlnum{1}\hlopt{:}\hlnum{6}\hlstd{]} \hlopt{~} \hlkwd{dmnorm}\hlstd{(mu_bm[}\hlnum{1}\hlopt{:}\hlnum{6}\hlstd{],} \hlkwc{cov} \hlstd{= var_bm[}\hlnum{1}\hlopt{:}\hlnum{6}\hlstd{,} \hlnum{1}\hlopt{:}\hlnum{6}\hlstd{])}
  \hlstd{\}}
  \hlstd{var_bm[}\hlnum{1}\hlopt{:}\hlnum{6}\hlstd{,} \hlnum{1}\hlopt{:}\hlnum{6}\hlstd{]} \hlkwb{<-} \hlnum{1}\hlopt{/}\hlstd{tau_bm} \hlopt{*} \hlkwd{diag}\hlstd{(}\hlkwd{rep}\hlstd{(}\hlnum{1}\hlstd{,} \hlnum{6}\hlstd{))}
  \hlstd{tau_bm} \hlopt{~} \hlkwd{dgamma}\hlstd{(}\hlnum{1}\hlstd{,} \hlnum{1}\hlstd{)}

  \hlkwa{for} \hlstd{(j} \hlkwa{in} \hlnum{1}\hlopt{:}\hlnum{6}\hlstd{) \{}
    \hlstd{mu_bm[j]} \hlopt{~} \hlkwd{dnorm}\hlstd{(}\hlnum{0}\hlstd{,} \hlnum{1}\hlstd{)}
  \hlstd{\}}
\end{alltt}
\end{kframe}
\end{knitrout}
{\setlength{\parindent}{0cm} Finally for the model, the stick breaking process
corresponding to Equations~\eqref{eq:pzi} and~\eqref{eq:breaking} is depicted.}
\begin{knitrout}
\definecolor{shadecolor}{rgb}{0.965, 0.965, 0.965}\color{fgcolor}\begin{kframe}
\begin{alltt}
  zlatent[1:M] <- \hlkwd{stick_breaking}(vlatent[1:(M - 1)])
  
  \hlkwd{for} (j in 1:(M - 1)) \{
    vlatent[j] ~ \hlkwd{dbeta}(1, alpha)
  \}
  
  alpha ~ \hlkwd{dgamma}(1, 1)
  tau_y ~ \hlkwd{dgamma}(1, 1)
  
\})
\end{alltt}
\end{kframe}
\end{knitrout}

{\setlength{\parindent}{0cm} With the full model defined, we next declare the
data list, which is made up of the response \code{Y}, the covariates
\code{X[,1] to \code{X[,6]}}, and the matrix of
distances \code{Dist}.
The constants in the model also need to be supplied, including the
number of locations \code{S}, the number of starting clusters \code{M}, and the
upper endpoint \code{D} for the uniform distribution of bandwidth. In addition,
the initial values are specified. Code to compile the model, supply the initial
values, and invoke the MCMC process is included in the supplementary package.}

\begin{knitrout}
\definecolor{shadecolor}{rgb}{0.965, 0.965, 0.965}\color{fgcolor}\begin{kframe}
\begin{alltt}
\hlstd{SLMMdata} \hlkwb{<-} \hlkwd{list}\hlstd{(}\hlkwc{y} \hlstd{= y,} \hlkwc{x1} \hlstd{= X[,}\hlnum{1}\hlstd{],} \hlkwc{x2} \hlstd{= X[,}\hlnum{2}\hlstd{],} \hlkwc{x3} \hlstd{= X[,}\hlnum{3}\hlstd{],}
                 \hlkwc{x4} \hlstd{= X[,}\hlnum{4}\hlstd{],} \hlkwc{x5} \hlstd{= X[,}\hlnum{5}\hlstd{],} \hlkwc{x6} \hlstd{= X[,}\hlnum{6}\hlstd{],}
                 \hlkwc{Dist} \hlstd{= distmatrix)}

\hlstd{SLMMConsts} \hlkwb{<-} \hlkwd{list}\hlstd{(}\hlkwc{S} \hlstd{=} \hlnum{159}\hlstd{,} \hlkwc{M} \hlstd{=} \hlnum{50}\hlstd{,} \hlkwc{D} \hlstd{=} \hlnum{100}\hlstd{)}

\hlstd{SLMMInits} \hlkwb{<-} \hlkwd{list}\hlstd{(}\hlkwc{tau_y} \hlstd{=} \hlnum{1}\hlstd{,}
                  \hlkwc{latent} \hlstd{=} \hlkwd{rep}\hlstd{(}\hlnum{1}\hlstd{, SLMMConsts}\hlopt{$}\hlstd{S),} \hlkwc{alpha} \hlstd{=} \hlnum{2}\hlstd{,}
                  \hlkwc{tau_bm} \hlstd{=} \hlnum{1}\hlstd{,}
                  \hlkwc{mu_bm} \hlstd{=} \hlkwd{rnorm}\hlstd{(}\hlnum{6}\hlstd{),} \hlkwc{phi} \hlstd{=} \hlnum{1}\hlstd{,} \hlkwc{tau_w} \hlstd{=} \hlnum{1}\hlstd{,}
                  \hlkwc{vlatent} \hlstd{=} \hlkwd{rbeta}\hlstd{(SLMMConsts}\hlopt{$}\hlstd{M} \hlopt{-} \hlnum{1}\hlstd{,} \hlnum{1}\hlstd{,} \hlnum{1}\hlstd{)}
\hlstd{)}
\end{alltt}
\end{kframe}
\end{knitrout}

\subsection{Inference of MCMC results}\label{sec:sum_mcmc}

The estimated parameters, together with the cluster assignments $\bm{z}$, are
determined for each replicate from the best post burn-in iteration selected
using the Dahl's method \citep[][]{dahl2006model}.
\cite{dahl2006model}
proposed a least-square model-based clustering for estimating the clustering of
observations using draws from a posterior clustering distribution. In this
method, membership matrices for each iteration, $\bm{B}^{(1)},\ldots,\bm{B}^{(M)}$,
where $M$ is the number of post-burn-in MCMC itertations, are calculated.
The membership matrix for the $c$th iteration, $\bm{B}^{(c)}$ is defined as:
\begin{align}
\bm{B}^{(c)} = (\bm{B}^{(c)}(i,j))_{i,j\in \{1:n\}} = 1(z_i^{(c)} = z_j^{(c)})_{n\times
n},
\end{align}
with $1()$ being the indicator function, $\bm{B}^{(c)}(i,j) \in \{0,1\}$ for all
$i,j = 1,...,n$ and $c=1,\ldots, M$. Having $\bm{B}^{(c)}(i,j)=1$ means
observations $i$ and $j$ are in the same cluster in the $c$th iteration. The
average of $\bm{B}^{(1)},\ldots, \bm{B}^{(M)}$ can be calculated as
\begin{equation*}
	\bar{\bm{B}} = \frac{1}{M} \sum_{c=1}^M \bm{B}^{(c)},
\end{equation*}
where $\sum$ here denotes element-wise summation of matrices. The $(i,j)$th
entry of $\bar{\bm{B}}$ provides an empirical estimate of the probability for
locations $i$ and $j$ to be in the same cluster.

Next we find the iteration that has the least squared distance
to $\bar{\bm{B}}$ as:
\begin{align}
C_{LS} = \argmin_{c \in (1:M)} \sum_{i=1}^n \sum_{j=1}^n 
\left(\bm{B}^{(c)}(i,j) -
\bar{\bm{B}}(i,j)\right)^2,
\end{align}
where $\bm{B}^{(c)}(i,j)$ is the ($i,j$)th entry of $\bm{B}^{(c)}$, and
$\bar{\bm{B}}(i,j)$ is the ($i,j$)th entry of $\bar{\bm{B}}$. An advantage of
the least-squares clustering is the fact that information from all clusterings
are utilized via the usage of the empirical pairwise probability matrix
$\bar{\bm{B}}$. It is also intuitively appealing, as the average clustering is
selected instead of formed via an external, \textit{ad hoc} clustering
algorithm.

\subsection{Model Assessment}

In the spatial regression model, the Gaussian process spatial
structure~\blue{$\bm{\Sigma_W} = \sigma_w^2 \bm{H}$}
can be constructed via several different weighting schemes including the
aforementioned unity, exponential, and Gaussian schemes in
\eqref{eq:weightingscheme}. In order to determine which weighting scheme is the
most suitable for the data, a commonly used model comparison criterion, the
logarithm of the pseudo-Marginal likelihood
\citep[LPML;][]{ibrahim2013bayesian}, is applied. The LPML can be obtained
through the conditional predictive ordinate (CPO) values. With $Y^*_{(-i)} =
(Y_1,\ldots, Y_{i-1}, Y_{i+1},\ldots, Y_n)$ denoting the observations with
the~$i$th subject response deleted, CPO can be regarded as
leave-one-out-cross-validation under Bayesian framework, and it estimates the
probability of observing~$Y_i$ in the future if after having already observed
$Y^{*}_{(-i)}$. The CPO for the~$i$th subject is calculated as:
\begin{equation}
	\label{eq:CPO}
\text{CPO}_i = \int
f(y(s_i)\mid\bm{\beta}(s_i),w(s_i),\sigma_y^2)\pi(\bm{w}(\bm{s}),
\bm{\beta}(\bm{s}), \sigma_y^2\mid Y^*_{(-i)}) \dd (\bm{w}(\bm{s}),
\bm{\beta}(\bm{s}),
\sigma_y^2),
\end{equation} 
where 
\begin{equation*}
\pi(\bm{w}(\bm{s}), \bm{\beta}(\bm{s}), \sigma_y^2\mid Y^*_{(-i)}) =
\frac{\prod_{j \ne i}
f(y(s_j)\mid \bm{\beta}(s_j),w(s_j),\sigma_y^2)\pi(\bm{w}(\bm{s}),
\bm{\beta}(\bm{s}),
\sigma_y^2\mid Y^*_{(-i)})}{c(Y^*_{(-i)})},
\end{equation*}
and $c(Y^*_{(-i)})$ is the normalizing constant. Within the Bayesian framework,
a
Monte Carlo estimate of the CPO can be obtained as:
\begin{equation}
	\label{eq:CPOest}
\widehat{\text{CPO}}_i^{-1} = \frac{1}{M} \sum_{t=1}^{M}
\frac{1}{f(y(s_i)\mid w_t(s_i), \bm{\beta}_t(s_i), \sigma_{yt}^2)},
\end{equation}
where $w_t(s_i)$ is calculated based on the sampled $\phi$ in the $t$-th
iteration, and $\bm{\beta}_t(s_i)$ and~$\sigma_{yt}^2$ are, respectively, the
$t$-th iteration samples for~$\bm{\beta}(s_i)$ and~$\sigma_y^2$. An estimate of
the LPML can subsequently be calculated as:
\begin{equation}
	\label{eq:LPML}
	\widehat{\text{LPML}} = \sum_{i=1}^{N} \log(\widehat{\text{CPO}}_i).
\end{equation}
A model with a larger LPML value is preferred. In addition, $p_D$, the
effective number of parameters, which can be used to measure the complexity of
the model, is defined as
\begin{gather} \label{eq:pD}
	p_D = \bar{D} - D(\bar{\boldsymbol{\theta}}),
\end{gather}
where $D = -2 \log f(\bm{y}(\bm{s})\mid\boldsymbol{\theta})$ is the deviance of
the model,
$\bar{D}$ is the posterior mean of deviance, $\bar{\boldsymbol{\theta}}$ is the
posterior mean of the parameters and $D(\bar{\boldsymbol{\theta}})$ denotes
deviance at posterior means.

\subsection{Convergence Diagnostics}
We use the Rand index \citep[RI;][]{rand1971objective} to measure the accuracy
of clustering. The RI is defined as 
\begin{eqnarray*}
 \mathrm{RI} = (a+b)/(a+b+c+d) = (a+b)\left.\right/{n \choose 2 }, 
\end{eqnarray*}
where $\mathcal{C}_1 = \{ X_1, \ldots, X_r \}$ and $\mathcal{C}_2 = \{Y_1,
\ldots, Y_s\}$ are two partitions of $\{ 1, 2, \ldots, n\}$, and $a, b, c$ and
$d$ respectively denote the number of pairs of elements of $\{1, 2, \ldots, n\}$
that are (a) in a same set in $\mathcal{C}_1$ and a same set in $\mathcal{C}_2$,
(b) in different sets in $\mathcal{C}_1$ and different sets in $\mathcal{C}_2$,
(c) in a same set in $\mathcal{C}_1$ but in different sets in $\mathcal{C}_2$,
and (d) in different sets in $\mathcal{C}_1$ and a same set in $\mathcal{C}_2$.
The RI ranges from 0 to 1 with a higher value indicating better agreement
between the two partitions. In particular, $\mathrm{RI} = 1$ indicates that
$\mathcal{C}_1$ and $\mathcal{C}_2$ are identical in terms of modulo labeling of
the nodes.

\section{Simulation Studies}\label{sec:simu}

In this section, we conduct simulation studies to assess the performance of the
proposed methods under scenarios where there is no clustered covariate effect,
and when there is indeed clustered covariate effect. All simulations are
run on an institutional high performance computing cluster running
Red Hat Enterprise Linux Server (release 6.7).

\subsection{Simulation Without Clustered Covariate Effects}\label{ssec:simnull}

The spatial adjacency structure of counties in Georgia is used. As a starting
point, to mimic the real dataset \blue{we use later}, one observation is
generated for each of the 159 counties. Six covariate vectors are generated for
the 159 counties with each entry i.i.d. from $N(0,1)$, making a $159\times 6$
covariate matrix $\bm{X}$. The spatial random effects~$\bm{w}$ are simulated
based on the matrix of great circle distance (GCD) between county centroids. The
great circle distances are obtained using the function \textsf{distCosine()},
and the centroids are calculated based on county polygons using the function
\textsf{centroid()}, both provided by the \textsf{R} package \textbf{geosphere}
\citep[][]{Rpkg:geosphere}. The GCD matrix is subsequently normalized to have a
maximum value of 10 for ease in computation. and the response vector~$\bm{Y}$ is
generated as
\begin{equation*}
  \bm{Y} = \bm{X}\bm{\beta} + \bm{w} + \bm{\epsilon},
\end{equation*}
where $\bm{X} = (X_1,\ldots, X_6)$, $\bm{w}\sim \text{MVN}(\bm{0},
\exp(-\mbox{GCD}/4))$, and $\bm{\epsilon} \sim
\mbox{MVN}(\bm{0},\bm{I})$. Different values of $\bm{\beta}$ are used:
$(1, 0, 1, 0, 0.5, 2)^\top$, $(2, 0, 1, 0, 4, 2)^\top$, and $(9, 0, -4, 0, 2,
5)^\top$, corresponding to scenarios where the signal is weak, moderate, and
strong. For each of the three $\bm{\beta}$'s, the average parameter estimate
denoted by $\bar{\hat{\beta}}_{\ell,m}$ ($\ell = 1,\cdots,159$; $m=1,\cdots, 6$) in
100 simulations is calculated as
\begin{gather}
\bar{\hat{\beta}}_{\ell,m} = \frac{1}{100} \sum_{r=1}^{100}
\hat{\beta}_{\ell,m,r},
	\label{eq:betahat}
\end{gather}
where $\hat{\beta}_{\ell,m,r}$ denotes the posterior estimate for the~$m$th
coefficient of county~$\ell$ in the~$r$th replicate. The performance of these
posterior estimates are evaluated by the mean absolute bias (MAB), the mean
standard deviation (MSD), the mean of mean squared error (MMSE) and mean
coverage rate (MCR) of the 95\% highest posterior density (HPD) intervals in the
following ways:
\begin{align}
\text{MAB} &= \frac{1}{159} \sum_{\ell=1}^{159} \frac{1}{100} \sum_{r=1}^{100}
\left|\hat{\beta}_{\ell,m,r} - \beta_{\ell,m} \right|,\label{eq:MAB}\\
\text{MSD} &= \frac{1}{159} \sum_{\ell=1}^{159}
\sqrt{\frac{1}{99}
\sum_{r=1}^{100} \left(\hat{\beta}_{\ell,m,r} - \bar{\hat{\beta}}_{\ell,m}
\right)^2},\label{eq:MSD}\\
\text{MMSE}& =  \frac{1}{159} \sum_{\ell=1}^{159} \frac{1}{100}
\sum_{r=1}^{100}
\left(\hat{\beta}_{\ell,m,r} - \beta_{\ell,m} \right)^2,\label{eq:MMSE}\\
\text{MCR} &=  \frac{1}{159} \sum_{\ell=1}^{159} \frac{1}{100}
\sum_{r=1}^{100}
1 \left( \hat{\beta}_{\ell,m,r} \in \text{95\% HPD interval}
\right)\label{eq:MCR},
\end{align}

In each replicate, the MCMC chain length is set to be 50,000, with thinning 10
and the first 2,000 samples are discarded as burn-in, therefore we have 3,000
samples for posterior inference. The parameter $D$ for the uniform prior of
bandwidth, i.e. $\phi$ in Equation~\eqref{eq:weightingscheme}, is set to 100
such that the prior for bandwidth is also
noninformative. In Table~\ref{tab:simnull} the average
parameter estimates $\bar{\hat{\beta}}_{\ell,m}$ are reported together with the
four performance measures in Equations~\eqref{eq:MAB},\eqref{eq:MSD},
\eqref{eq:MMSE} and \eqref{eq:MCR} are reported for the three settings. Under
all three settings, the parameter estimates are highly close to the true
underlying values, and have very small MAB, MSD and MMSE, while maintaining the
MCR at close to 95\% level. The RI's are all very close to or equal to 1,
indicating that the clustering results are highly consistent and credible. It is
worth noticing that, even when the signal is relatively weak, the clustering
approach is quite precise.

\begin{table}[tbp]
	\centering
	\caption{Average parameter estimates and performance of parameter
		estimates
and clustering results when without clustered covariate
effect.}\label{tab:simnull}
	\begin{tabular}{cccccccc}
		\toprule
		& & $\bar{\hat{\beta}}$ &MAB & MSD & MMSE & MCR & RI  \\
		\midrule 
Setting 1 & $\beta_1$ & 1.005 & 0.110 & 0.085 & 0.007 & 0.910 & 0.997 \\
   & $\beta_2$ & 0.002 & 0.087 & 0.069 & 0.005 & 0.980 &  \\
   & $\beta_3$ & 0.996 & 0.104 & 0.076 & 0.006 & 0.950 &  \\
   & $\beta_4$ & -0.003 & 0.090 & 0.071 & 0.005 & 0.970 &  \\
   & $\beta_5$ & 0.508 & 0.106 & 0.080 & 0.006 & 0.960 &  \\
   & $\beta_6$ & 1.978 & 0.109 & 0.081 & 0.007 & 0.970 &  \\
[.5ex]
Setting 2 & $\beta_1$ & 1.971 & 0.218 & 0.117 & 0.014 & 0.930 & 0.999 \\
   & $\beta_2$ & -0.001 & 0.090 & 0.072 & 0.005 & 0.980 &  \\
   & $\beta_3$ & 0.987 & 0.129 & 0.085 & 0.007 & 0.960 &  \\
   & $\beta_4$ & 0.002 & 0.098 & 0.076 & 0.006 & 0.950 &  \\
   & $\beta_5$ & 3.934 & 0.362 & 0.145 & 0.021 & 0.910 &  \\
   & $\beta_6$ & 1.955 & 0.194 & 0.105 & 0.011 & 0.960 &  \\[.5ex]
Setting 3 & $\beta_1$ & 9.006 &  0.108 & 0.081 & 0.007 & 0.930 & 1.000 \\
   & $\beta_2$ & -0.001 & 0.088 & 0.069 & 0.005 & 0.970 &  \\
   & $\beta_3$ & -3.996 & 0.089 & 0.069 & 0.005 & 0.960 &  \\
   & $\beta_4$ & -0.002 & 0.091 & 0.072 & 0.005 & 0.960 &  \\
   & $\beta_5$ & 2.007 & 0.106 & 0.081 & 0.007 & 0.950 &  \\
   & $\beta_6$ & 4.990 & 0.090 & 0.071 & 0.005 & 0.960 &  \\
		\bottomrule
	\end{tabular}
\end{table}

\begin{figure}[tbp]
	\centering
	\includegraphics[width = \textwidth]{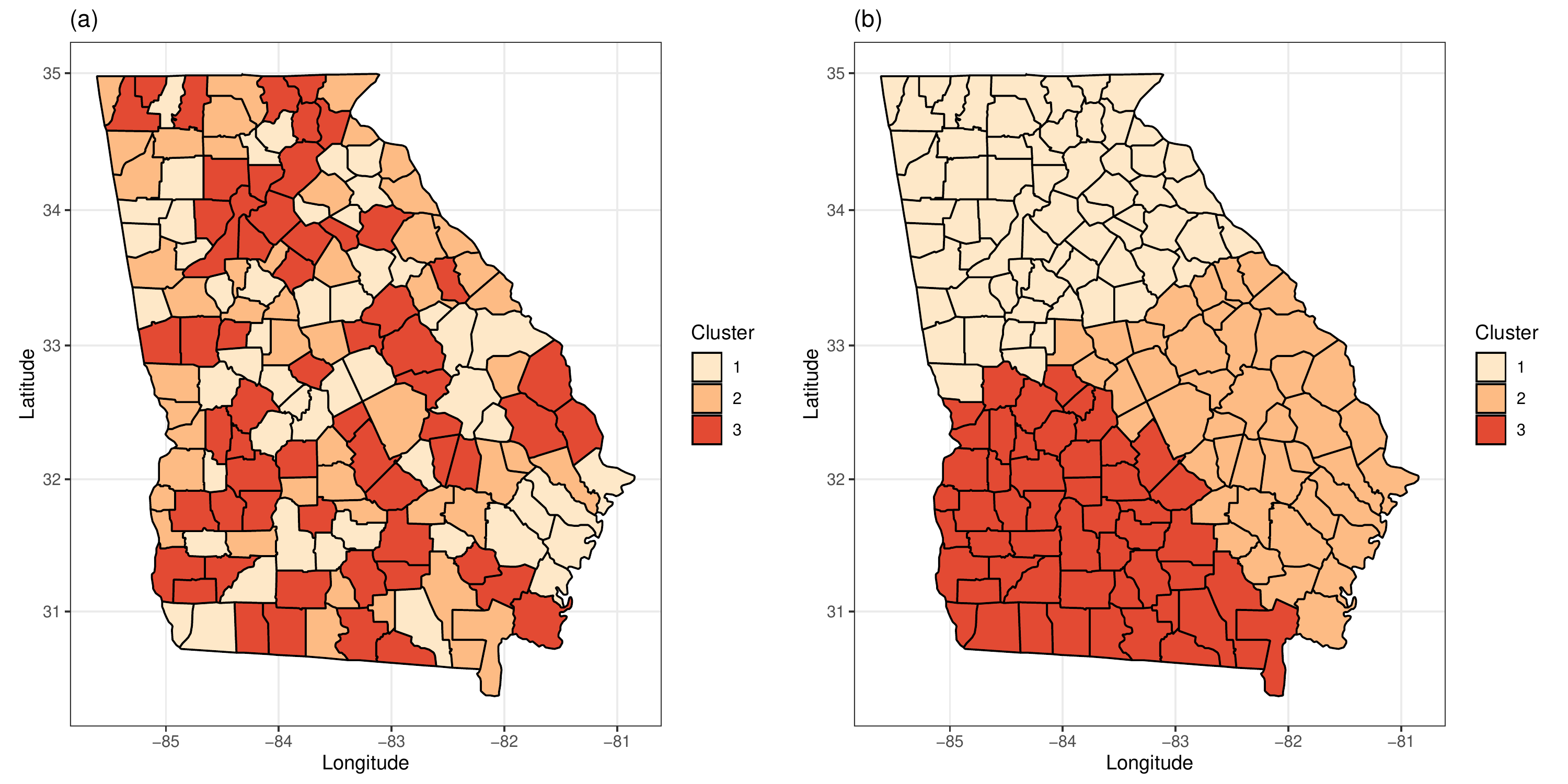}
	\caption{Visualization of (a) random cluster assignment and (b) regional
cluster assignment for Georgia counties used for simulation studies.}
	\label{fig:GArandom}
\end{figure}

\subsection{Simulation with Clustered Covariate Effects}

We consider an underlying setting where there exist clustered covariate effects.
First we consider a setting where the clustered covariate effect is independent
of spatial locations, i.e. where cluster belonging are set randomly. The 159
counties are randomly assigned to three clusters, visualized in
Figure~\ref{fig:GArandom}(a). There are, respectively, 51, 49, and 59 counties
in the three clusters. Different parameter vectors are used for data generation
in different clusters (see Table~\ref{tab:regionpars}) to assess the estimation
and clustering performance under different strengths of signals. The spatial
random effect $\bm{W}$ is generated using the same setting as before. The
performance measures are presented in Table~\ref{tab:simrand}. In another
scenario, a setting where the clustered covariate effect depends on spatial
locations. Consider a partition of Georgia counties into three large regions,
visualized in Figure~\ref{fig:GArandom}(b). The same parameter vectors in
Table~\ref{tab:regionpars} are used for the three clusters under three settings.
Corresponding performance measures are reported in Table~\ref{tab:simregion}.

For each signal strength and each of the two settings, we randomly selected four
replicates from the total of 100 replicates and visualize the results in the
Online Supplement. It is no surprise that under both settings, the accuracy of
clustering increases with the strengthening of signals. It can be seen from
Supplemental~Figures~1 and~4 that with weak true signals, the proposed approach
suffers from over-clustering, which is a known property of Dirichlet process
mixtures that the posterior does not concentrate at the true number of clusters
\citep[see, e.g.,][]{miller2013simple}. This over-clustering behavior, however,
diminishes as the signals' strength increase. When the signals are strong, the
RI reaches near 0.85, indicating that 85\% of the time, two counties that belong
to the same cluster are correctly put into the same cluster. Together with
increase in RI is decrease in MCR, which is an inevitable result of
incorporating more counties in each cluster. For each county, taking other
counties that do not belong to this county's true cluster introduces bias in
estimation.

\begin{table}[tbp]
	\centering
	\caption{True parameter vectors used in data generation
for three clusters.}\label{tab:regionpars}
\begin{tabular}{cllll}
	\toprule
	 & Cluster 1 & Cluster 2 & Cluster 3 \\
	 \midrule 
	 Setting 1 & (1, 0, 1, 0, 0.5, 2) & (1, 0.7, 0.3, 2, 0, 3) & (2, 1, 0.8, 1, 0,
1)
\\
	 Setting 2 & (2, 0, 1, 0, 4, 2) & (1, 0, 3, 2, 0, 3) & (4, 1, 0, 3, 0, 1)\\
	 Setting 3 & (9, 0, -4, 0, 2, 5) & (1, 7, 3, 6, 0, -1) & (2, 0, 6, 1, 7, 0) \\
	 \bottomrule
	\end{tabular}
\end{table}

\begin{table}[tbp]
	\centering
	\caption{Performance of parameter
		estimates
and clustering results under the scenario where cluster
belongings are set randomly.}\label{tab:simrand}
	\begin{tabular}{ccccccc}
		\toprule
		& &MAB & MSD & MMSE & MCR & RI  \\
		\midrule 
Setting 1 & $\beta_1$ & 0.186 & 0.421 & 0.242 & 0.935 & 0.621 \\
   & $\beta_2$ & 0.173 & 0.401 & 0.180 & 0.967 &  \\
   & $\beta_3$ & 0.150 & 0.293 & 0.093 & 0.985 &  \\
   & $\beta_4$ & 0.206 & 0.721 & 0.672 & 0.924 &  \\
   & $\beta_5$ & 0.168 & 0.241 & 0.063 & 0.977 &  \\
   & $\beta_6$ & 0.227 & 0.747 & 0.705 & 0.916 &  \\ [.5ex]
Setting 2 & $\beta_1$ & 0.967 & 0.812 & 0.735 & 0.757 & 0.690 \\
   & $\beta_2$ & 0.443 & 0.378 & 0.150 & 0.834 &  \\
   & $\beta_3$ & 0.958 & 0.806 & 0.743 & 0.753 &  \\
   & $\beta_4$ & 0.961 & 0.826 & 0.698 & 0.762 &  \\
   & $\beta_5$ & 1.390 & 1.071 & 1.339 & 0.786 &  \\
   & $\beta_6$ & 0.670 & 0.608 & 0.408 & 0.766 &  \\ [.5ex]
Setting 3 & $\beta_1$ & 1.941 & 0.763 & 0.636 & 0.816 & 0.852 \\
   & $\beta_2$ & 1.870 & 0.728 & 0.602 & 0.867 &  \\
   & $\beta_3$ & 2.310 & 0.933 & 0.940 & 0.828 &  \\
   & $\beta_4$ & 1.491 & 0.636 & 0.445 & 0.814 &  \\
   & $\beta_5$ & 1.700 & 0.738 & 0.606 & 0.845 &  \\
   & $\beta_6$ & 1.442 & 0.600 & 0.388 & 0.828 &  \\
		\bottomrule
	\end{tabular}
\end{table}

\begin{table}[tbp]
	\centering
	\caption{Performance of parameter
		estimates
and clustering results under the scenario where cluster
belongings are set depending on the county centroid
locations.}\label{tab:simregion}
	\begin{tabular}{ccccccc}
		\toprule
		& &MAB & MSD & MMSE & MCR & RI  \\
		\midrule 
Setting 1 & $\beta_1$ & 0.201 & 0.413 & 0.237 & 0.966 & 0.597 \\
   & $\beta_2$ & 0.154 & 0.426 & 0.195 & 0.977 &  \\
   & $\beta_3$ & 0.165 & 0.289 & 0.091 & 0.991 &  \\
   & $\beta_4$ & 0.216 & 0.720 & 0.657 & 0.937 &  \\
   & $\beta_5$ & 0.182 & 0.256 & 0.068 & 0.988 &  \\
   & $\beta_6$ & 0.210 & 0.683 & 0.619 & 0.904 &  \\ [.5ex]
Setting 2 & $\beta_1$ & 0.885 & 0.806 & 0.782 & 0.787 & 0.691 \\
   & $\beta_2$ & 0.425 & 0.353 & 0.138 & 0.860 &  \\
   & $\beta_3$ & 0.789 & 0.783 & 0.832 & 0.802 &  \\
   & $\beta_4$ & 0.969 & 0.856 & 0.758 & 0.783 &  \\
   & $\beta_5$ & 1.437 & 1.106 & 1.280 & 0.789 &  \\
   & $\beta_6$ & 0.585 & 0.592 & 0.432 & 0.807 &  \\ [.5ex]
Setting 3 & $\beta_1$ & 1.981 & 0.764 & 0.625 & 0.831 & 0.855 \\
   & $\beta_2$ & 1.711 & 0.706 & 0.710 & 0.855 &  \\
   & $\beta_3$ & 2.389 & 0.951 & 0.983 & 0.838 &  \\
   & $\beta_4$ & 1.363 & 0.609 & 0.486 & 0.852 &  \\
   & $\beta_5$ & 1.617 & 0.724 & 0.634 & 0.830 &  \\
   & $\beta_6$ & 1.471 & 0.603 & 0.389 & 0.838 &  \\
		\bottomrule
	\end{tabular}
\end{table}

\section{Real Data Analysis}\label{sec:real_data}

\begin{figure}[tbp]
  \centering 
  \includegraphics[width=\textwidth]{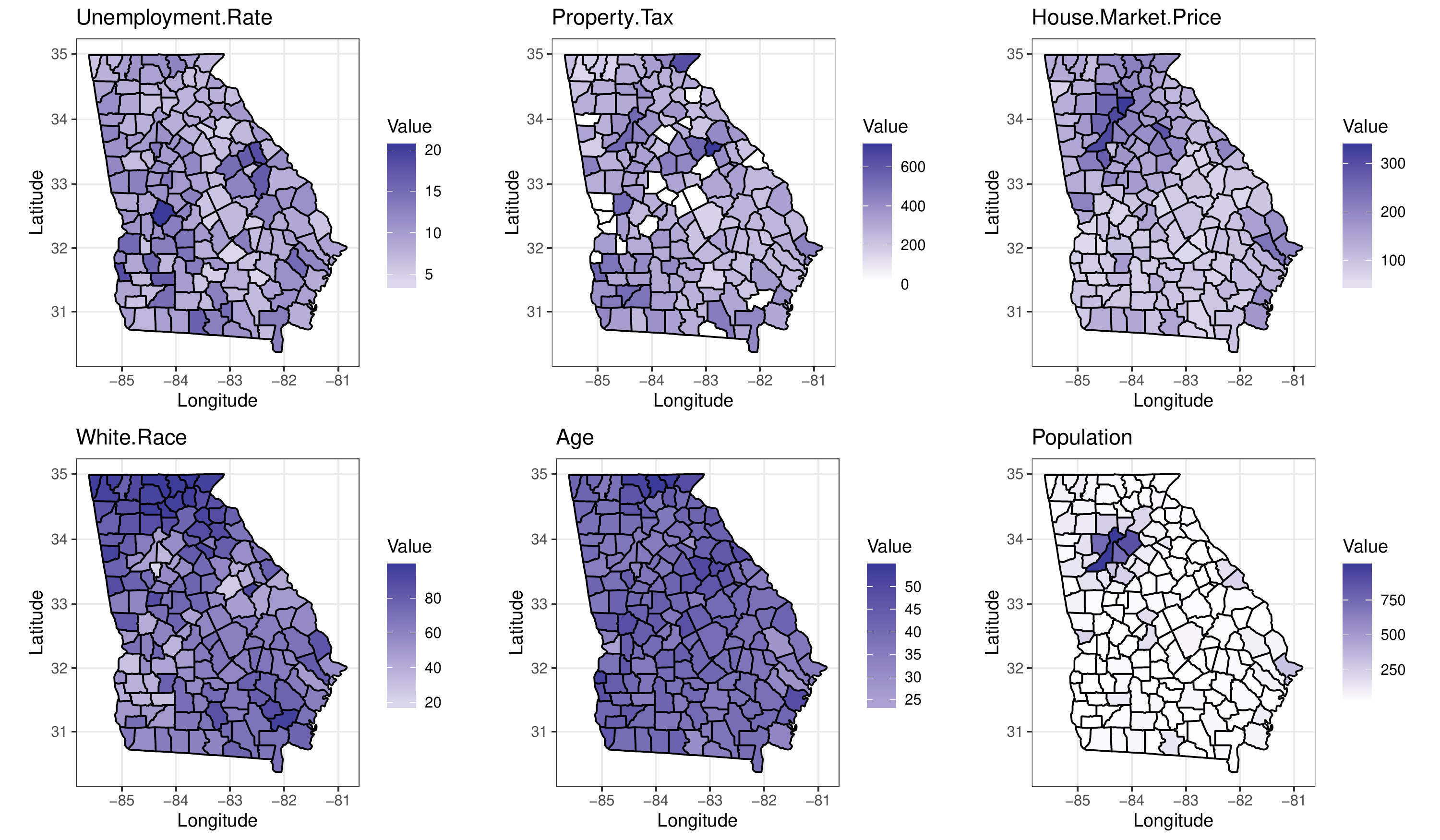}
  \caption{Visualizations of covariate values in counties of Georgia.}
  \label{fig:varplot}
\end{figure}

We consider analyzing influential factors for monthly housing cost in Georgia
using the proposed methods. The dataset is available at
\url{www.healthanalytics.gatech.edu}, with 159 observations corresponding to the
159 counties in Georgia. For each county, the dependent variable median monthly
housing cost for all occupied housing units is observed. The independent
variables considered here include: the percentage of adults aged 18 to 64 who
are unemployed ($X_1$), the average total real and personal property taxes
collected per person ($X_2$), the median home market value ($X_3$, in thousand
dollars), the percentage of White race population ($X_4$), the median age
($X_5$), and size of a county's population ($X_6$, in thousands).
Figure~\ref{fig:varplot} provides a visualization of the 6 covariates on the
Georgia map. In the computation, the covariates are centered and scaled to have
mean~0 and unit standard deviation. Also, following the common practice in
economics to account for long-tailed distributions, we take the logarithm of
monthly housing cost before fitting the model. The response variable is also
centered and scaled, and therefore all models to follow are fitted without the
intercept term.

We firstly apply the model assessment criteria, LPML, for selecting the best
weighting scheme for the data. The LPML values for the unity weighting scheme,
the exponential weighting scheme and the Gaussian weighting scheme are shown in
Table~\ref{tab:LPML}. From Table~\ref{tab:LPML} we can see that the model with
exponential weighting scheme has the largest LPML value among the three
candidate schemes, and is therefore preferred.
To verify that there is indeed spatially varying covariate effects, we also
fitted the spatially-varying coefficients model without
clustering~\eqref{hierarchical model} to the dataset. The model is also compared
against a vanilla Bayesian regression, \blue{where no spatial effect is
considered, and observations are treated as i.i.d. samples from the population.
In this model, no spatial variation is assumed in the covariate
effects~$\bm{\beta}$, and the model reduces to the regular Bayesian linear
regression.}

The computation is performed on a desktop computer running Windows 10
Enterprise, with i7-8700K CPU @ 3.70GHz. The computing time as well as
performance measure for these three models are recorded and presented in
Table~\ref{tab:lpmlmodels}. The proposed model takes around 650 seconds to run,
followed by the spatially varying coefficients model, and then the spatially
constant coefficients model. The LPML values of the first two models that allow
for spatially varying coefficients are larger than the vanilla regression model,
and the differences are not minor. This indicates that there indeed exist
spatially varying covariate effect, and more flexible models are preferred.
Comparing the LPML and $p_D$ for the first two models, it can be seen that the
proposed model reduces $p_D$ and provides better fit to the data. Combining the
conclusions from Tables~\ref{tab:LPML} and~\ref{tab:lpmlmodels}, the proposed
model with exponential weighting scheme for $\bm{W}(s)$ is fitted on the
dataset.

\begin{table}[tbp]
	\centering
	\caption{LPML values for different weighting
schemes in the proposed model.}\label{tab:LPML}
	\begin{tabular}{lcccc}
		\toprule
		& Unity & Exponential & Gaussian & \\
		\midrule 
		LPML & -165.784
		 & -165.620 & -217.878
		 \\
		\bottomrule
	\end{tabular}
\end{table} 

\begin{table}
\centering 
\caption{LPML and $p_D$ values and computing time
for different models.}\label{tab:lpmlmodels}
\begin{tabular}{lccc}
\toprule 
& The proposed model & Spatial-varying coefficients model & Vanilla regression\\
\midrule 
LPML & -165.620 & -174.171 & -203.013\\ 
$p_D$ & 75.17 & 105.92 & 11.44\\
Time (seconds) & 653.00 & 318.24 & 260.55\\
\bottomrule
\end{tabular}
\end{table}

\begin{figure}[tbp]
    \centering
    \includegraphics[width = 0.6\textwidth]{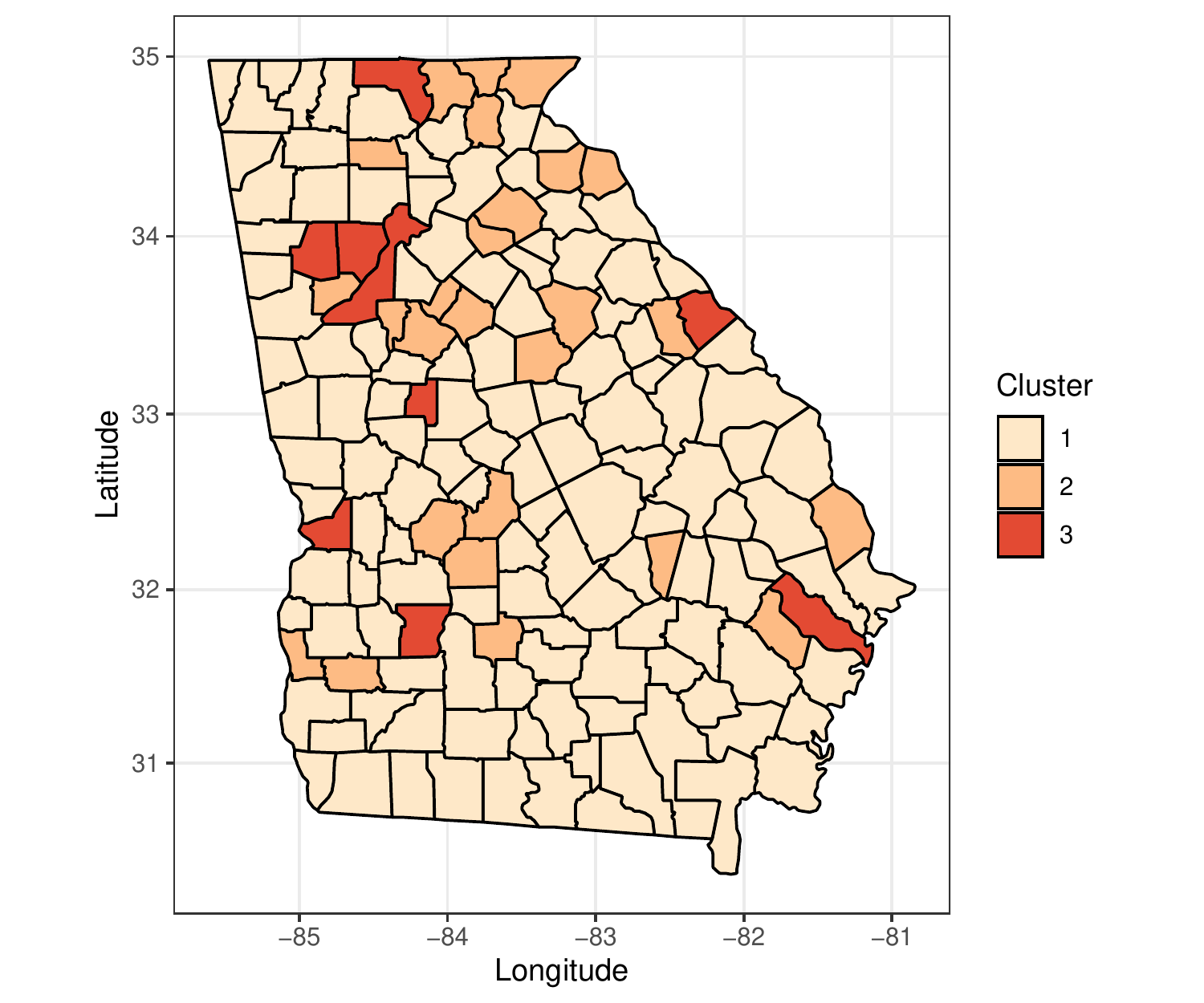}
    \caption{Clusters produced by the proposed approach.}
    \label{fig:clusters}
\end{figure}

A total of 3 clusters are identified. The cluster belongings of the 159 counties
are visualized in Figure~\ref{fig:clusters}, and their corresponding parameter
estimates are presented in Table~\ref{tab:parest}. From
Figure~\ref{fig:clusters} we can find that the cluster distribution is more
similar with the spatial distributions of the covariates population size and
median home market price. These two covariates also show great impact on
cluster~1, the largest cluster we obtained from the model. For cluster~1, which
includes most of the counties (124 out of 159), higher unemployment rates,
higher median home market value, and larger population sizes are significant
indicators of higher housing costs. For cluster~2 (26 out of 159), median home
market value is also positively correlated with the monthly housing cost, while
higher median age indicates lower housing cost. For cluster~3 (9 out of 159)
median home market value turns out to be the only decisive factor and has
significant increasing effect for housing cost. These results indicate that for
most counties of Georgia, unemployment rates, median home market value and
population size drive the variation of housing costs greatly. However, not all
the counties have the same pattern. Housing costs of some counties are affected
by median home market price and median age instead, and for a few counties, the
housing costs are related to median home market value instead of the other
covariates. This example here verifies the fact that the proposed model can
detect the spatial clusters which share similar covariate effects.

\begin{table}
  \centering
\caption{Parameter estimates and their 95\% HPD intervals for the three clusters
identified.}\label{tab:parest}
  \begin{tabular}{cccc}
  \toprule 
      Cluster &1 & 2 & 3  \\ \midrule 
	  $\hat{\beta}_1$ &  0.154 & -0.088 & -0.349 \\
	  & (0.020, 0.293) & (-0.424, 0.394) & (-1.533, 0.811)  \\
	 $\hat{\beta}_2$ & -0.091 & 0.043 & -0.047\\
	  & (-0.217, 0.040) & (-0.404, 0.433) & (-1.19, 1.08)  \\
	 $\hat{\beta}_3$ & 1.841 & 1.985 &  1.540\\
	  & (1.615, 2.072) & (1.463, 2.580) & (0.317, 2.778)\\
	 $\hat{\beta}_4$ & -0.078 &  -0.146 &  0.151\\
	  & (-0.295, 0.102) & (-0.648, 0.313) & (-0.910, 1.509) \\
	 $\hat{\beta}_5$ &  -0.004 & -0.998 & -0.994 \\
	  & (-0.205, 0.192) & (-1.478, -0.578) & (-1.899, 0.186)  \\
	 $\hat{\beta}_6$ & 0.327 & 0.611 & 0.119  \\
	  & (0.069, 0.698) & (-0.592, 1.421) & (-0.789, 1.435) \\
\bottomrule
  \end{tabular}
\end{table}

\section{Discussion}\label{sec:discussion}

In this paper, we have proposed a Bayesian clustered coefficients linear
regression model with spatial random effects to capture heterogeneity of
regression coefficients. Multiple weighting schemes in modeling the spatial
random effects have been proposed, and the corresponding Bayesian model
selection criterion have been discussed. Compared to a vanilla regression model
with no spatial random effect, allowing the covariate effects to be spatially
varying provides better fit to the data, and more profound insight into
heterogeneity in development at different locations. In addition, compared to
observations made in \cite{ma2019bayesian}, where each location is allowed to
have its own set of parameter estimates, the clustering approach reduces the
effective number of parameters without sacrificing the model goodness-of-fit.
The usage of the method is illustrated both in simulation studies and an
application to analysis of impacting factors for housing cost in Georgia.

A few topics beyond the scope of this paper are worth further investigation. In
this paper, we only considered the full model that includes all covariates.
Appropriate approaches for variable selection under a clustered regression
context is worth investigating. The DPMM is used to get clustering information
of regression coefficients. The posterior on the number of clusters is not
consistent based on the DPMM. Such pattern have been observed in both our
simulation studies, where there are some small clusters which only contain a few
counties. Proposing a consistent prior
\citep{geng2019probabilistic,hu2020bayesian} for clustered regression
coefficients is an important future work. In addition, extending our approach in
non-gaussian model is an interesting topic. Considering spatial dependent
structure for the regression coefficients \citep{zhao2020bayesian} is devoted to
future research.

\end{document}